\documentclass[twocolumn,showpacs,amsmath,amssymb,aps,prl]{revtex4-1}
\usepackage{graphicx,color}
\usepackage[bookmarks=false,colorlinks=true,citecolor=blue,linkcolor=blue,menucolor=blue,urlcolor=blue]{hyperref}

\begin{document}

\title{Role of particle-number statistics
in interference of independent Bose fields}
\author{Toru Kawakubo}
\author{Katsuji Yamamoto}
\affiliation{Department of Nuclear Engineering, Kyoto University,
Kyoto 606-8501, Japan}
\date{\today}

\begin{abstract}
We elucidate generally the interference of independent Bose fields
in view of the conditional probability for the particle number measurements,
and clarify its relation to the source number statistics.
Despite lack of intrinsic phases, the interference phase can be inferred
from the particle number registered at one detector by using
the classical mean fields.  If the conditional number distributions
for the other detectors, given the outcome of the first detector,
exhibit sufficiently narrow peaks around the values specified
by the estimated phases, the mean field description is valid
in a single run of interference.  The widths in the conditional distribution
are determined by the number statistics of the sources, among which
notable scaling behavior is found depending on the detector configurations
with the boundary at the Poissonian.  The mean field description
is found to be applicable to Poissonian and sub-Poissonian sources,
whereas for super-Poissonian sources it is likely invalidated
with the rather broad conditional distribution.
\end{abstract}


\maketitle

Interference is often considered as a signature of superposition
in quantum systems. In particular, interference in many-body systems
as a macroscopic quantum effect has been attracting many interests.
In usual experiments, two states originating from a common source
are subject to interfere, namely, each particle interferes with itself.
However, in many-boson systems including lasers
\cite{Magyar1963, Pfleegor1967} and atomic Bose-Einstein condensates (BECs)
\cite{Andrews1997}, interference between independently prepared particles
has also been observed.  Such interference is often explained
with the spontaneous symmetry breaking for the relative phase,
which gives nonvanishing expectation values of the field operators
or mean fields.  In BECs, a U(1) symmetry is relevant for
the global phase rotation of atomic wavefunctions, the breakdown of which
relies on a nonphysical interaction \cite{Naraschewski1996, Leggett2006}.
In optical systems, a U(1) symmetry also arises from lack of
an absolute phase reference, which is ensured by
the effective photon-number conservation in optical processes
\cite{Molmer1997, Sanders2003, Bartlett2007}.  The U(1) symmetry breaking
hence seems problematic in the absence of real mechanism.

The interference observed for independent sources under the U(1) symmetry
has been attributed to the back-action of particle detection on the systems,
which causes localization of the relative phase in a single run
\cite{Javanainen1996, Molmer1997, Sanders2003}.
Another approach to the interference is to calculate
the correlation functions of the particle numbers measured by
the different detectors, which show the spatial modulation.
By evaluating the statistical moments of the Fourier components of
the spatial modulation up to the fourth order, the plane-wave interference
of atomic BECs is predicted in a single run with a random phase
\cite{Naraschewski1996, Iazzi2011}.
This analysis exploits the nature of the plane-wave mode functions.
Generally, some common understanding will be presented
for the interference appearing under various configurations,
which is based on the probability theory on quantum measurement.
Moreover, there will be some intimate relationship between the interference
and the particle-number statistics of sources,
by considering the fact that the interference is observed
so far for lasers (Poissonian states) and BECs (sub-Poissonian states).

In this paper, we investigate the interference of independent Bose fields
under general configurations for sources and detectors,
and clarify its relation to the source statistics.
We examine the joint probabilities of the particle numbers registered
by the detectors directly, rather than the correlations,
to see the interference in a single run.
The outcome at one detector provides information about the relative phase,
despite lack of intrinsic phases due to the U(1) symmetry.  This information
appears in the conditional distributions for the particle numbers
at the other detectors, which are derived from the joint probabilities with
the given outcome of the first detector.  The relative phase is estimated
by applying the mean field description to the measurement outcome.
If the conditional distribution has sufficiently narrow peaks around
the values predicted by the estimated phases, it is almost certain
that the outcome at the second detector takes a value close to
one of the mean field predictions.  Hence, the conditional distribution
provides a quantitative criterion for the validity of
the mean field description.  The mean field description is found
to be applicable to Poissonian and sub-Poissonian sources,
whereas for super-Poissonian sources it is likely invalidated
with the rather broad conditional distribution.

We consider a system of noninteracting Bose particles,
photons or cold atoms, where two independent sources are contained.
The positive-frequency field operator $ \hat{\psi}(\mathbf{x}, t) $
is given generally in terms of the annihilation operators $ \hat{a}_l $
for a complete set of mode functions $ \{ \phi_l \} $:
$ \hat{\psi}(\mathbf{x},t) = \sum_l \hat{a}_l \phi_l (\mathbf{x},t) $, where
the time evolution of the free field is represented in the mode functions
$ \phi_l (\mathbf{x},t) $, which is determined in practice by expanding
$ \hat{\psi} $ alternatively in terms of the plane-wave modes.
In order to describe an interference experiment, the mode functions
are chosen suitably to provide the two independent sources
as $ \hat{a}_1 \equiv \hat{a} $ and $ \hat{a}_2 \equiv \hat{b} $.
For example, in interference between two wavepackets of light
the wavevector distributions are localized around the central wavevectors of
the respective sources.  In the case of two atomic BECs \cite{Andrews1997},
the initial mode functions $ \phi_l (\mathbf{x},0)$ are divided into two groups
consisting of the eigenstates of the respective one-particle Hamiltonians
with harmonic traps.  In the following we assume for simplicity
that all the particles are populated in the two source modes ($ l = 1, 2 $),
while the other modes ($ l \geq 3 $) are in the vacuum states.
(This will be almost valid in typical interference experiments.)
Then, the density matrix for the sources is given
by $ \hat{\rho} = \hat{\rho}_a \otimes \hat{\rho}_b $,
where each source state, respecting the U(1) symmetry, is given
with the particle-number statistics $ p_s (N) $ \cite{Sanders2003} as
\begin{align}
\hat{\rho}_s &= \sum_{N=0}^\infty p_s (N)|N \rangle \langle N| &( s = a, b ).
\label{eq:rho-s}
\end{align}

In the photon measurement for optical interference experiments,
a commonly used photodetector records the number of photoelectrons
emitted from the detector surface during a time interval $ T $.
The time and surface integrated photon-flux operator for the photoelectron
emission at the detector $m$ is given 
\cite{Kelley1964,*Cook1982,*Bondurant1985}
by
\begin{equation}
\hat{I}_m = \eta_m \int_0^T dt \int_{S_m} dxdy\,
\hat{\psi}^\dagger(\mathbf{x},t) \hat{\psi}(\mathbf{x},t) ,
\label{eq:I_m}
\end{equation}
where $ \eta_m $ is the quantum efficiency, and the $ z $ axis is taken
normal to the detector surface $S_m$.  The bandwidth $ \Delta \omega $
of the incident radiation is assumed to be small enough compared
with the central frequency $ \omega_0 $.  The photon-flux operators
in Eq.\ (\ref{eq:I_m}) are specifically expressed
as bilinear forms of the mode operators,
$ \hat{I}_m = \sum_{ll'} R^{(m)}_{ll'} \hat{a}_l^\dagger \hat{a}_{l'} $,
with the Hermitian matrices $ R^{(m)} $ obtained from Eq.\ (\ref{eq:I_m})
by substitution $ \hat{\psi}^\dagger \hat{\psi} \to \phi_l^* \phi_{l'} $.
For the detection of cold atoms, we may take a resonant interaction
between the atomic internal levels and the probe light, which transfers
the information of the atomic density to $\hat{\psi}^\dagger \hat{\psi} $
of the probe light \cite{Goldstein1998}.  Hence, the detection of cold atoms
is treated in the same way as the photon number detection.

The joint probabilities of the photon counts $ n_1, \dotsc, n_M $
by the $M$ detectors ($ 1 \leq M \leq M_\text{end} $),
which characterize the full statistics of interference, are given by
\begin{equation}
P(n_1, \dotsc, n_M) = \left\langle \mathopen{:} \prod_{m=1}^M \frac{1}{n_m!}
( \hat{I}_m )^{n_m} e^{-\hat{I}_m} \mathclose{:} \right\rangle ,
\label{eq:PM}
\end{equation}
where $ \mathopen{:} \mathclose{:} $ stands for normal ordering
\cite{Kelley1964,*Cook1982,*Bondurant1985}.
The flux operators are presented explicitly as
\begin{equation}
\hat{I}_m = R^{(m)}_{aa} \hat{a}^\dagger\hat{a}
+ R^{(m)}_{bb}\hat{b}^\dagger\hat{b}
+ R^{(m)}_{ab}\hat{a}^\dagger\hat{b} + R^{(m)}_{ba}\hat{b}^\dagger\hat{a} .
\label{eq:I_m3}
\end{equation}
Here, it should be noted that the terms involving the vacuum modes
($ l \geq 3 $) are dropped in $ \hat{I}_m $
since they provide null contributions to Eq.\ (\ref{eq:PM})
as the normal-ordered expectation values.
The mean particle number measured at each detector is given by
\begin{equation}
\langle n_m \rangle = \langle \hat{I}_m \rangle
= R^{(m)}_{aa} \bar{N}_a + R^{(m)}_{bb} \bar{N}_b .
\label{eq:n_m}
\end{equation}
Here, $\bar{N}_s = \operatorname{Tr} [ \hat{\rho}_s \hat{s}^\dagger \hat{s} ] $
are the mean particle numbers initially contained in the sources, which are
assumed to be large enough to produce $ \langle n_m \rangle \gg 1 $ for
high accuracy statistics. The coefficients $R^{(m)}_{aa}$ and $R^{(m)}_{bb}$
indicate the probabilities for each particle from the respective sources
to fall into the detector $ m $.  They may represent the resolution of
interference.  Specifically, $ R^{(m)}_{ss} \propto 1/M_\text{end} \to 0 $,
but keeping $R^{(m)}_{ss}\bar{N}_s \gg 1$ for $ \langle n_m \rangle \gg 1 $,
when the particles are measured by almost continuously distributed
many detectors, resulting in a fine interference pattern,
e.g., spatial interference fringes \cite{Magyar1963,Andrews1997}.

In the above sense, as seen in Eq.\ (\ref{eq:n_m}),
a change of $R^{(m)}_{ss}$ (or resolution) for the detectors
may be viewed alternatively as an modification of the source statistics.
Here, consider scaling of the detector matrices
(by removing several detectors and changing the quantum efficiencies),
\begin{align}
\tilde{R}^{(m)}(q;M) &= R^{(m)} / q & (m = 1, \dotsc , M)
\label{eq:scaling}
\end{align}
with $ \tilde{R}^{(m)}(q;M) = 0 $ ($m > M$),
and define the binomial distribution
\begin{align}
B^{N}_{N'} (q) &\equiv \binom{N}{N'} q^{N'} (1-q)^{N-N'} &(0 \leq N' \leq N) .
\label{eq:Bq}
\end{align}
In evaluating the joint probabilities, $ \langle \mathopen{:}
( \hat{I}_1 )^{k_1} \dotsm ( \hat{I}_M )^{k_M} \mathclose{:} \rangle $
contained in Eq.\ (\ref{eq:PM}) are calculated for a Fock state
$ | N_a , N_b \rangle $ with the normal-ordered expectation values
$ \langle (\hat{b}^\dagger)^{k_b} (\hat{a}^\dagger)^{k_a}
\hat{a}^{k_a} \hat{b}^{k_b} \rangle
= [N_a !/(N_a-k_a)!] \times [N_b !/(N_b-k_b)!] $
($ k_a + k_b = k_1 + \dotsb + k_M $),
which are multiplied by $ q^{k_a} q^{k_b} $ under the scaling.
Then, by considering the relation $ q^k [N!/(N-k)!]
= \sum_{N^\prime = k}^N B^{N}_{N^\prime} (q) [N^\prime !/(N^\prime - k)!] $,
the effects of this scaling can be renormalized to the source statistics
without changing the calculations in Eq.\ (\ref{eq:PM}) as
\begin{equation}
\tilde{p}_s (N;q) = \sum_{N'=N}^\infty p_s (N') B^{N'}_{N} (q) ,
\label{eq:eff_s}
\end{equation}
which is also normalized as the original $ p_s (N) $.
Hence, the number statistics of the sources
may be replaced with the effective ones in Eq.\ (\ref{eq:eff_s})
for any scaling of $ q $, reproducing the same joint probability
for the measurement by the $ M $ detectors (namely the $M$-detector model):
\begin{equation}
\{ \tilde{R}^{(m)}(q;M) , \tilde{p}_s (N;q) \} \to P(n_1, \dotsc, n_M) .
\label{eq:R-p-P}
\end{equation}
This may be viewed as a renormalization transformation
among the number statistics.  It indicates universal relation
for various interference phenomena, ranging from two-mode homodyne detection
($ M=2 $) to measurement of spacial fringes ($ M=M_\text{end} \gg 1 $).
According to Eq.\ (\ref{eq:eff_s}), the mean $ \tilde{\bar{N}}_s $ and
variance $ \tilde{V}_s $ for the effective statistics are given
in terms of the original ones as $ \tilde{\bar{N}}_s = q \bar{N}_s $
and $ \tilde{V}_s = q^2 V_s + (1-q)q \bar{N}_s $.  Then,
for a sub-Poissonian distribution ($ V_s < \bar{N}_s $), the effective one
is still sub-Poissonian ($ \tilde{V}_s < \tilde{\bar{N}}_s $) as
\begin{equation}
\tilde{V}_s / \tilde{\bar{N}}_s = q ( V_s / \bar{N}_s ) + 1 - q .
\label{eq:V-N-eff}
\end{equation}
The Poissonian form is preserved under the renormalization
up to the scaling of mean as $ \tilde{\bar{N}}_s = q \bar{N}_s $.
On the other hand, for a super-Poissonian distribution
the effective one is still super-Poissonian.

We now examine the validity of the mean field description
for interference phenomena, where the field operators are replaced
with c-numbers as $\hat{a} \to \alpha $ and $\hat{b} \to \beta$
(expectation values for coherent states $ | \alpha , \beta \rangle $).
Specifically, we have
\begin{align}
\bar{n}_m
&= \langle \alpha , \beta | \hat{I}_m | \alpha , \beta \rangle \nonumber \\
&= \langle n_m \rangle + 2|R^{(m)}_{ab}| \bar{N}_a^{1/2} \bar{N}_b^{1/2}
\cos( \delta_{ab} + \theta_m ) ,
\label{eq:mean}
\end{align}
where $ \bar{N}_a = | \alpha |^2 $, $ \bar{N}_b = | \beta |^2 $,
$ \delta_{ab} = \arg \alpha - \arg \beta $, $ \theta_m = \arg R^{(m)}_{ab}$,
and $ \langle n_m \rangle $ is the same as Eq.\ (\ref{eq:n_m})
for the U(1)-invariant sources.  The set of $ \{ \bar{n}_m \} $ exhibits
the interference pattern with the cosine term in Eq.\ (\ref{eq:mean}),
which oscillates with $\theta_m$ depending on the detector location.
The mean field description is, however, not directly applicable
to the U(1)-invariant sources in Eq.\ (\ref{eq:rho-s}) with
$\langle\hat{a}^\dagger\hat{b}\rangle = 0$, eliminating the cosine term
in  Eq.\ (\ref{eq:mean}).  Nevertheless, by experiments and
theoretical calculations the interference fringes are observed
in a single run with a random relative phase for Poissonian sources
(laser fields \cite{Magyar1963}) and sub-Poissonian sources
(optical number states \cite{Molmer1997, Sanders2003}
and BECs \cite{Andrews1997, Naraschewski1996, Iazzi2011}).

We hence consider the relationship between the interference phenomena and
the source number statistics.  Specifically, we examine the validity of
the mean field description by inspecting the joint probability $P(n_1,n_2)$
for any pair of detectors, say 1 and 2,
depending on the source statistics.  Given the outcome $n_1$ at detector 1,
the mean field description in Eq.\ (\ref{eq:mean}) provides an estimate
for the relative phase, generally with two possibilities
$ \delta_{ab}^{\pm} $ due to the cosine.
Then, the outcome $n_2$ at detector 2 is inferred with the estimated phases:
\begin{equation}
\bar{n}_1 = n_1 \rightarrow \delta_{ab}^{\pm} (n_1)
\rightarrow \bar{n}_2 [ \delta_{ab}^{\pm} (n_1) ] .
\end{equation}
If the actual count $ n_2 $ is close to one of
$ \bar{n}_2 [ \delta_{ab}^{\pm} (n_1) ] $, fixing the estimation of
$ \delta_{ab}$, we find that the interference occurs as described by
the mean (classical) fields.  This criterion for the interference
can be checked readily by calculating the conditional distribution
$P_\text{c}(n_2|n_1)$ from $ P(n_1 , n_2) $ with given $n_1$.
If $P_\text{c}(n_2|n_1)$ has sufficiently narrow peaks
at $ \bar{n}_2 [ \delta_{ab}^{\pm} (n_1) ] $, the second outcome $n_2$
should be close to either of the peaks with high probability.
Specifically, the width of the peak should be no greater than
that of the Poisson distribution $ e^{-\bar{n}_2}(\bar{n}_2)^{n_2}/{n_2}! $,
which is the shot noise level for the coherent states
$| \alpha, \beta \rangle $.  Here, we conjecture that sub-Poissonian sources
lead to the narrow peaks, showing the interference pattern.
It is pointed out \cite{Pegg2009} that wavepackets
emitted from a cavity maintain a pronounced relative phase coherence
when the intracavity field has a narrow number distribution.
Light beams from such sub-Poissonian cavities will exhibit the interference.
This phase coherence of each source is essential
to fix the interference phase in the number measurements.

Consider first the case of fine detector resolution with $| R^{(m)} | \ll 1$
in the usual measurement of spatial interference fringes.
This case can be treated by scaling as the two-detector model
with $ \tilde{R}^{(1,2)} = R^{(1,2)}/q \sim 1 $ and $ q \to 0 $,
which provides the same $ P(n_1 , n_2) $ with the effective statistics
in Eq.\ (\ref{eq:R-p-P}).  Then, as seen in Eq.\ (\ref{eq:V-N-eff}),
the effective statistics of sub-Poissonian sources
approach the Poissonian for $ q \to 0 $.
Hence, by using any sub-Poissonian sources, essentially the same result
is obtained for the interference fringes as the Poissonian case,
where the mean field description is valid as numerically confirmed
in the following.  This is not the case for super-Poissonian sources.
For $ R^{(m)} = q \tilde{R}^{(m)} \to 0 $ with $ \tilde{R}^{(m)} \sim 1$
fixed, the large $ \bar{N}_s = \tilde{\bar{N}}_s / q \propto 1/|R^{(m)}| $,
which is required to produce $ \langle n_m \rangle \gg 1 $, may
derive even the larger $ V_s $, e.g., $ V_s \propto \bar{N}_s^2 $,
for a super-Poissonian source, giving a nonzero
$ q ( V_s / \bar{N}_s ) $ for $ q \to 0 $ in Eq.\ (\ref{eq:V-N-eff}).

In order to examine the validity of the mean field description
for general $ R^{(m)}_{ss}$, we have calculated numerically
$P_\text{c}(n_2|n_1)$ by using Eq.\ (\ref{eq:PM}) for some typical sources.
The detector matrices are chosen for instance
as $ R^{(1)}_{aa} = R^{(2)}_{bb} = 0.6 R $,
$ R^{(1)}_{bb} = R^{(2)}_{aa} = 0.4 R $,
$ |R^{(1,2)}_{ab}|^2 = R^{(1,2)}_{aa} R^{(1,2)}_{bb} $,
giving the maximum interference term in Eq.\ (\ref{eq:mean}),
with the relative phase $ \theta_2 - \theta_1 = 0.9 \pi $.
The first outcome is set as $n_1 = 118$,
which corresponds to $ \delta_{ab}^\pm (n_1) + \theta_1 = \mp 1.39 $
and $\bar{n}_2[\delta_{ab}^\pm (n_1)] \approx 53, 113 $.
Due to limitation on the numerical calculation,
$ R \bar{N}_a = R \bar{N}_b = 100 $ are taken, giving
$ \langle n_1 \rangle = \langle n_2 \rangle = 100 $ with
$ R^{(1,2)}_{aa} + R^{(1,2)}_{bb} = R $, and consistently
$ \langle n_1 \rangle + \langle n_2 \rangle = 200 \approx 118+(53+113)/2 $.
The scaling for the effective statistics is also used by taking
$ R/q = \tilde{R} = 0.867 $ to calculate $P_\text{c}(n_2|n_1)$
for the increasing $ \bar{N}_{a,b} = 100/R $
with the smaller $ R $, after it is checked numerically
for $ q \sim 0.5 $ with $ \bar{N}_{a,b} = 25/R $.
A bound on $ \theta_2 - \theta_1 $ may appear for the increasing $ R $
from the condition $ \bar{n}_1 + \bar{n}_2 \leq \bar{N}_a + \bar{N}_a $
($ = 200/R $) due to the unitarity or the total number conservation,
e.g., $ 0.9 \pi \leq \theta_2 - \theta_1 \leq \pi $ for $ R = 0.867 $.
This is clearly seen in the familiar two-mode homodyne detection,
where $ \hat{I}_{1,2} =( \hat{a}^\dagger \pm \hat{b}^\dagger )
( \hat{a} \pm \hat{b} )/2 $ with $ e^{i(\theta_2 - \theta_1)} = -1 $.

\begin{figure}[t]
\includegraphics{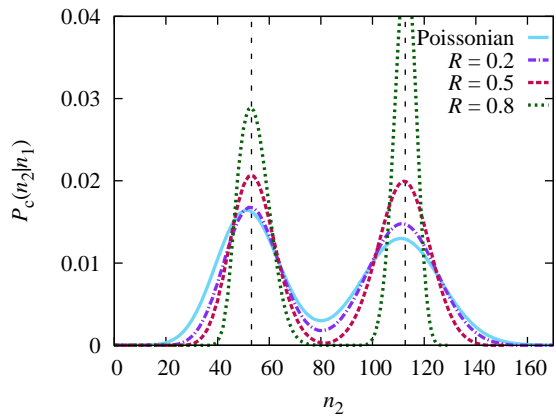}
\caption{(Color online)
Conditional distribution for number states
$| N/R, N/R \rangle$ with $ N = 100$ and some values of $R$.
The Poissonian case corresponds to $R \to 0$.
The mean field values $\bar{n}_2[\delta_{ab}^\pm (n_1)]
\approx 53, 113 $ for $n_1 = 118$ are indicated with vertical dotted lines.}
\label{fig:NN}
\end{figure}

The results for number states $ | N/R, N/R \rangle $ with $ N = 100$
and some values of $ R $ are shown in Fig.\ \ref{fig:NN}.
The case of Poissonian source is also plotted for comparison,
corresponding to $ R \to 0 $,
where the Poisson distribution $ \propto ( \bar{n}_2 )^{n_2}/{n_2}! $
for $ n_2 $ is confirmed around the peaks
(though rather broad due to not so large $ \langle n_{1,2} \rangle = 100 $).
The peaks agree with $\bar{n}_2[\delta_{ab}^\pm (n_1)] \approx 53, 113 $
(vertical dotted lines), and exhibit the narrower widths
than the Poissonian case.  Therefore, the mean field description is valid
for these sub-Poissonian number states and also their effective statistics,
i.e., the binomial distributions in Eq.\ (\ref{eq:Bq}).
Here, the limit $ R \to 1 $ becomes unphysical
with the dominating $\bar{n}_2[\delta_{ab}^+ (n_1)] $
to give $ n_1 + n_2 \approx 118+113 > 200 (R=1) $, violating the unitarity.

\begin{figure}[t]
\includegraphics{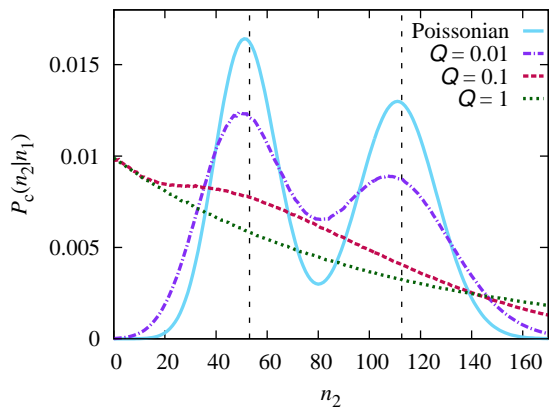}
\caption{(Color online)
Conditional distribution for super-Poissonian sources
with $\mathcal{P}(\alpha)$ in Eq.\ (\ref{eq:Palpha}).
The mean field values are shown the same as in Fig.\ \ref{fig:NN}.}
\label{fig:Palpha}
\end{figure}
We have also considered a super-Poissonian source
with a U(1)-invariant $ \mathcal{P} $-representation as
\begin{equation}
\mathcal{P}(\alpha) \propto (| \alpha |^2/Q\bar{N})^{1/Q-1}
\exp (-| \alpha |^2/ Q\bar{N}) ,
\label{eq:Palpha}
\end{equation}
where $ Q = (V - \bar{N})/\bar{N}^2 $ with $ Q > 0 $.
The limit $Q \to 0$ corresponds to the Poissonian, whereas $Q = 1$ to
the thermal state.  The conditional distribution is shown
in Fig.\ \ref{fig:Palpha}, which does not depend on $R$ in this case
with $ R \bar{N}_{a,b} $ (=100) fixed.  The increasing $Q$ broadens
the distribution, eventually washing out the peaks.
We have further examined the single-photon-added thermal state 
\cite{Agarwal1992,*Zavatta2007}.
This nonclassical super-Poissonian state has the variance smaller than
the thermal case.  Despite this fact, for the small $ R $
($ R \bar{N}_{a,b} $ fixed), the conditional distribution becomes flatter
than that for the thermal sources.  These results indicate that
the behavior of interference is rather complicated for
super-Poissonian sources, likely invalidating the mean field description.

To conclude, in view of the conditional probability
for the number measurements, we have elucidated the common mechanism
for the interference of independent Bose fields under various situations,
ranging from two-mode homodyne interference to spacial fringes.
The interference is determined by the source number statistics,
among which the scaling behavior is present depending on
the detector characteristics with the boundary at the Poissonian.
For sub-Poissonian and Poissonian sources the interference pattern appears
in a single run, consistently with the mean field description,
whereas this is not the case for super-Poissonian sources.
It will be a challenge for future experiments
to confirm the role of source statistics with the scaling behavior,
by preparing various source states and detector configurations.

T. K. was supported by the JSPS Grant No.\ 22.1355.

%


\begin{thebibliography}{17}%
\makeatletter
\providecommand \@ifxundefined [1]{%
 \@ifx{#1\undefined}
}%
\providecommand \@ifnum [1]{%
 \ifnum #1\expandafter \@firstoftwo
 \else \expandafter \@secondoftwo
 \fi
}%
\providecommand \@ifx [1]{%
 \ifx #1\expandafter \@firstoftwo
 \else \expandafter \@secondoftwo
 \fi
}%
\providecommand \natexlab [1]{#1}%
\providecommand \enquote  [1]{``#1''}%
\providecommand \bibnamefont  [1]{#1}%
\providecommand \bibfnamefont [1]{#1}%
\providecommand \citenamefont [1]{#1}%
\providecommand \href@noop [0]{\@secondoftwo}%
\providecommand \href [0]{\begingroup \@sanitize@url \@href}%
\providecommand \@href[1]{\@@startlink{#1}\@@href}%
\providecommand \@@href[1]{\endgroup#1\@@endlink}%
\providecommand \@sanitize@url [0]{\catcode `\\12\catcode `\$12\catcode
  `\&12\catcode `\#12\catcode `\^12\catcode `\_12\catcode `\%12\relax}%
\providecommand \@@startlink[1]{}%
\providecommand \@@endlink[0]{}%
\providecommand \url  [0]{\begingroup\@sanitize@url \@url }%
\providecommand \@url [1]{\endgroup\@href {#1}{\urlprefix }}%
\providecommand \urlprefix  [0]{URL }%
\providecommand \Eprint [0]{\href }%
\providecommand \doibase [0]{http://dx.doi.org/}%
\providecommand \selectlanguage [0]{\@gobble}%
\providecommand \bibinfo  [0]{\@secondoftwo}%
\providecommand \bibfield  [0]{\@secondoftwo}%
\providecommand \translation [1]{[#1]}%
\providecommand \BibitemOpen [0]{}%
\providecommand \bibitemStop [0]{}%
\providecommand \bibitemNoStop [0]{.\EOS\space}%
\providecommand \EOS [0]{\spacefactor3000\relax}%
\providecommand \BibitemShut  [1]{\csname bibitem#1\endcsname}%
\let\auto@bib@innerbib\@empty
\bibitem [{\citenamefont {Magyar}\ and\ \citenamefont
  {Mandel}(1963)}]{Magyar1963}%
  \BibitemOpen
  \bibfield  {author} {\bibinfo {author} {\bibfnamefont {G.}~\bibnamefont
  {Magyar}}\ and\ \bibinfo {author} {\bibfnamefont {L.}~\bibnamefont
  {Mandel}},\ }\href {\doibase 10.1038/198255a0} {\bibfield  {journal}
  {\bibinfo  {journal} {Nature}\ }\textbf {\bibinfo {volume} {198}},\ \bibinfo
  {pages} {255} (\bibinfo {year} {1963})}\BibitemShut {NoStop}%
\bibitem [{\citenamefont {Pfleegor}\ and\ \citenamefont
  {Mandel}(1967)}]{Pfleegor1967}%
  \BibitemOpen
  \bibfield  {author} {\bibinfo {author} {\bibfnamefont {R.~L.}\ \bibnamefont
  {Pfleegor}}\ and\ \bibinfo {author} {\bibfnamefont {L.}~\bibnamefont
  {Mandel}},\ }\href {\doibase 10.1103/PhysRev.159.1084} {\bibfield  {journal}
  {\bibinfo  {journal} {Phys. Rev.}\ }\textbf {\bibinfo {volume} {159}},\
  \bibinfo {pages} {1084} (\bibinfo {year} {1967})}\BibitemShut {NoStop}%
\bibitem [{\citenamefont {Andrews}\ \emph {et~al.}(1997)\citenamefont
  {Andrews}, \citenamefont {Townsend}, \citenamefont {Miesner}, \citenamefont
  {Durfee}, \citenamefont {Kurn},\ and\ \citenamefont
  {Ketterle}}]{Andrews1997}%
  \BibitemOpen
  \bibfield  {author} {\bibinfo {author} {\bibfnamefont {M.~R.}\ \bibnamefont
  {Andrews}}, \bibinfo {author} {\bibfnamefont {C.~G.}\ \bibnamefont
  {Townsend}}, \bibinfo {author} {\bibfnamefont {H.-J.}\ \bibnamefont
  {Miesner}}, \bibinfo {author} {\bibfnamefont {D.~S.}\ \bibnamefont {Durfee}},
  \bibinfo {author} {\bibfnamefont {D.~M.}\ \bibnamefont {Kurn}}, \ and\
  \bibinfo {author} {\bibfnamefont {W.}~\bibnamefont {Ketterle}},\ }\href
  {\doibase 10.1126/science.275.5300.637} {\bibfield  {journal} {\bibinfo
  {journal} {Science}\ }\textbf {\bibinfo {volume} {275}},\ \bibinfo {pages}
  {637} (\bibinfo {year} {1997})}\BibitemShut {NoStop}%
\bibitem [{\citenamefont {Naraschewski}\ \emph {et~al.}(1996)\citenamefont
  {Naraschewski}, \citenamefont {Wallis}, \citenamefont {Schenzle},
  \citenamefont {Cirac},\ and\ \citenamefont {Zoller}}]{Naraschewski1996}%
  \BibitemOpen
  \bibfield  {author} {\bibinfo {author} {\bibfnamefont {M.}~\bibnamefont
  {Naraschewski}}, \bibinfo {author} {\bibfnamefont {H.}~\bibnamefont
  {Wallis}}, \bibinfo {author} {\bibfnamefont {A.}~\bibnamefont {Schenzle}},
  \bibinfo {author} {\bibfnamefont {J.~I.}\ \bibnamefont {Cirac}}, \ and\
  \bibinfo {author} {\bibfnamefont {P.}~\bibnamefont {Zoller}},\ }\href
  {\doibase 10.1103/PhysRevA.54.2185} {\bibfield  {journal} {\bibinfo
  {journal} {Phys. Rev. A}\ }\textbf {\bibinfo {volume} {54}},\ \bibinfo
  {pages} {2185} (\bibinfo {year} {1996})}\BibitemShut {NoStop}%
\bibitem [{\citenamefont {Leggett}(2006)}]{Leggett2006}%
  \BibitemOpen
  \bibfield  {author} {\bibinfo {author} {\bibfnamefont {A.~J.}\ \bibnamefont
  {Leggett}},\ }\href@noop {} {\emph {\bibinfo {title} {{Quantum liquids: Bose
  condensation and Cooper pairing in condensed-matter systems}}}}\ (\bibinfo
  {publisher} {Oxford University Press},\ \bibinfo {address} {New York},\
  \bibinfo {year} {2006})\BibitemShut {NoStop}%
\bibitem [{\citenamefont {M{\o}lmer}(1997)}]{Molmer1997}%
  \BibitemOpen
  \bibfield  {author} {\bibinfo {author} {\bibfnamefont {K.}~\bibnamefont
  {M{\o}lmer}},\ }\href {\doibase 10.1103/PhysRevA.55.3195} {\bibfield
  {journal} {\bibinfo  {journal} {Phys. Rev. A}\ }\textbf {\bibinfo {volume}
  {55}},\ \bibinfo {pages} {3195} (\bibinfo {year} {1997})}\BibitemShut
  {NoStop}%
\bibitem [{\citenamefont {Sanders}\ \emph {et~al.}(2003)\citenamefont
  {Sanders}, \citenamefont {Bartlett}, \citenamefont {Rudolph},\ and\
  \citenamefont {Knight}}]{Sanders2003}%
  \BibitemOpen
  \bibfield  {author} {\bibinfo {author} {\bibfnamefont {B.~C.}\ \bibnamefont
  {Sanders}}, \bibinfo {author} {\bibfnamefont {S.~D.}\ \bibnamefont
  {Bartlett}}, \bibinfo {author} {\bibfnamefont {T.}~\bibnamefont {Rudolph}}, \
  and\ \bibinfo {author} {\bibfnamefont {P.~L.}\ \bibnamefont {Knight}},\
  }\href {\doibase 10.1103/PhysRevA.68.042329} {\bibfield  {journal} {\bibinfo
  {journal} {Phys. Rev. A}\ }\textbf {\bibinfo {volume} {68}},\ \bibinfo
  {pages} {042329} (\bibinfo {year} {2003})}\BibitemShut {NoStop}%
\bibitem [{\citenamefont {Bartlett}\ \emph {et~al.}(2007)\citenamefont
  {Bartlett}, \citenamefont {Rudolph},\ and\ \citenamefont
  {Spekkens}}]{Bartlett2007}%
  \BibitemOpen
  \bibfield  {author} {\bibinfo {author} {\bibfnamefont {S.~D.}\ \bibnamefont
  {Bartlett}}, \bibinfo {author} {\bibfnamefont {T.}~\bibnamefont {Rudolph}}, \
  and\ \bibinfo {author} {\bibfnamefont {R.~W.}\ \bibnamefont {Spekkens}},\
  }\href {\doibase 10.1103/RevModPhys.79.555} {\bibfield  {journal} {\bibinfo
  {journal} {Rev. Mod. Phys.}\ }\textbf {\bibinfo {volume} {79}},\ \bibinfo
  {pages} {555} (\bibinfo {year} {2007})}\BibitemShut {NoStop}%
\bibitem [{\citenamefont {Javanainen}\ and\ \citenamefont
  {Yoo}(1996)}]{Javanainen1996}%
  \BibitemOpen
  \bibfield  {author} {\bibinfo {author} {\bibfnamefont {J.}~\bibnamefont
  {Javanainen}}\ and\ \bibinfo {author} {\bibfnamefont {S.~M.}\ \bibnamefont
  {Yoo}},\ }\href {\doibase 10.1103/PhysRevLett.76.161} {\bibfield  {journal}
  {\bibinfo  {journal} {Phys. Rev. Lett.}\ }\textbf {\bibinfo {volume} {76}},\
  \bibinfo {pages} {161} (\bibinfo {year} {1996})}\BibitemShut {NoStop}%
\bibitem [{\citenamefont {Iazzi}\ and\ \citenamefont
  {Yuasa}(2011)}]{Iazzi2011}%
  \BibitemOpen
  \bibfield  {author} {\bibinfo {author} {\bibfnamefont {M.}~\bibnamefont
  {Iazzi}}\ and\ \bibinfo {author} {\bibfnamefont {K.}~\bibnamefont {Yuasa}},\
  }\href {\doibase 10.1103/PhysRevA.83.033611} {\bibfield  {journal} {\bibinfo
  {journal} {Phys. Rev. A}\ }\textbf {\bibinfo {volume} {83}},\ \bibinfo
  {pages} {033611} (\bibinfo {year} {2011})}\BibitemShut {NoStop}%
\bibitem [{\citenamefont {Kelley}\ and\ \citenamefont
  {Kleiner}(1964)}]{Kelley1964}%
  \BibitemOpen
  \bibfield  {author} {\bibinfo {author} {\bibfnamefont {P.~L.}\ \bibnamefont
  {Kelley}}\ and\ \bibinfo {author} {\bibfnamefont {W.~H.}\ \bibnamefont
  {Kleiner}},\ }\href {\doibase 10.1103/PhysRev.136.A316} {\bibfield  {journal}
  {\bibinfo  {journal} {Phys. Rev.}\ }\textbf {\bibinfo {volume} {136}},\
  \bibinfo {pages} {A316} (\bibinfo {year} {1964})}\BibitemShut {NoStop}%
\bibitem [{\citenamefont {Cook}(1982)}]{Cook1982}%
  \BibitemOpen
  \bibfield  {author} {\bibinfo {author} {\bibfnamefont {R.~J.}\ \bibnamefont
  {Cook}},\ }\href {\doibase 10.1103/PhysRevA.25.2164} {\bibfield  {journal}
  {\bibinfo  {journal} {Phys. Rev. A}\ }\textbf {\bibinfo {volume} {25}},\
  \bibinfo {pages} {2164} (\bibinfo {year} {1982})}\BibitemShut {NoStop}%
\bibitem [{\citenamefont {Bondurant}(1985)}]{Bondurant1985}%
  \BibitemOpen
  \bibfield  {author} {\bibinfo {author} {\bibfnamefont {R.~S.}\ \bibnamefont
  {Bondurant}},\ }\href {\doibase 10.1103/PhysRevA.32.2797} {\bibfield
  {journal} {\bibinfo  {journal} {Phys. Rev. A}\ }\textbf {\bibinfo {volume}
  {32}},\ \bibinfo {pages} {2797} (\bibinfo {year} {1985})}\BibitemShut
  {NoStop}%
\bibitem [{\citenamefont {Goldstein}\ \emph {et~al.}(1998)\citenamefont
  {Goldstein}, \citenamefont {Zobay},\ and\ \citenamefont
  {Meystre}}]{Goldstein1998}%
  \BibitemOpen
  \bibfield  {author} {\bibinfo {author} {\bibfnamefont {E.~V.}\ \bibnamefont
  {Goldstein}}, \bibinfo {author} {\bibfnamefont {O.}~\bibnamefont {Zobay}}, \
  and\ \bibinfo {author} {\bibfnamefont {P.}~\bibnamefont {Meystre}},\ }\href
  {\doibase 10.1103/PhysRevA.58.2373} {\bibfield  {journal} {\bibinfo
  {journal} {Phys. Rev. A}\ }\textbf {\bibinfo {volume} {58}},\ \bibinfo
  {pages} {2373} (\bibinfo {year} {1998})}\BibitemShut {NoStop}%
\bibitem [{\citenamefont {Pegg}(2009)}]{Pegg2009}%
  \BibitemOpen
  \bibfield  {author} {\bibinfo {author} {\bibfnamefont {D.~T.}\ \bibnamefont
  {Pegg}},\ }\href {\doibase 10.1103/PhysRevA.79.053837} {\bibfield  {journal}
  {\bibinfo  {journal} {Phys. Rev. A}\ }\textbf {\bibinfo {volume} {79}},\
  \bibinfo {pages} {053837} (\bibinfo {year} {2009})}\BibitemShut {NoStop}%
\bibitem [{\citenamefont {Agarwal}\ and\ \citenamefont
  {Tara}(1992)}]{Agarwal1992}%
  \BibitemOpen
  \bibfield  {author} {\bibinfo {author} {\bibfnamefont {G.~S.}\ \bibnamefont
  {Agarwal}}\ and\ \bibinfo {author} {\bibfnamefont {K.}~\bibnamefont {Tara}},\
  }\href {\doibase 10.1103/PhysRevA.46.485} {\bibfield  {journal} {\bibinfo
  {journal} {Phys. Rev. A}\ }\textbf {\bibinfo {volume} {46}},\ \bibinfo
  {pages} {485} (\bibinfo {year} {1992})}\BibitemShut {NoStop}%
\bibitem [{\citenamefont {Zavatta}\ \emph {et~al.}(2007)\citenamefont
  {Zavatta}, \citenamefont {Parigi},\ and\ \citenamefont
  {Bellini}}]{Zavatta2007}%
  \BibitemOpen
  \bibfield  {author} {\bibinfo {author} {\bibfnamefont {A.}~\bibnamefont
  {Zavatta}}, \bibinfo {author} {\bibfnamefont {V.}~\bibnamefont {Parigi}}, \
  and\ \bibinfo {author} {\bibfnamefont {M.}~\bibnamefont {Bellini}},\ }\href
  {\doibase 10.1103/PhysRevA.75.052106} {\bibfield  {journal} {\bibinfo
  {journal} {Phys. Rev. A}\ }\textbf {\bibinfo {volume} {75}},\ \bibinfo
  {pages} {052106} (\bibinfo {year} {2007})}\BibitemShut {NoStop}%
\end{thebibliography}

\end{document}